\documentclass[11pt,twoside]{article}
\usepackage{asp2014}

\newcommand{\angstrom}{\mbox{\normalfont\AA}}

\aspSuppressVolSlug
\resetcounters

\begin{document}

\title{Feature Selection for Better Spectral Characterization or: How I Learned to Start Worrying and Love Ensembles}

\author{Sankalp~Gilda}
\affil{University of Florida, Gainesville, FL, U.S.A; \email{s.gilda@ufl.edu}}

\paperauthor{Sankalp~Gilda}{s.gilda@ufl.edu}{0000-0002-3645-4501
}{University of Florida}{Department of Astronomy}{Gainesville}{FL}{32611}{USA}


\begin{abstract}
An ever-looming threat to astronomical applications of machine learning is the danger of over-fitting data, also known as the `curse of dimensionality.' This occurs when there are fewer samples than the number of independent variables. In this work, we focus on the problem of stellar
parameterization from low-mid resolution spectra, with blended absorption lines. We address this problem using an iterative algorithm to sequentially prune redundant features from synthetic PHOENIX spectra, and arrive at an optimal set of wavelengths with the strongest correlation with each
of the output variables -- T$_{\rm eff}$, $\log g$, and [Fe/H]. We find that at any given resolution, most features (i.e., wavelength points) are not only redundant, but actually act as noise and decrease the accuracy of parameter retrieval.
\end{abstract}



\section{Introduction}
Technological developments in different domains have increased the amounts and complexity of data at an unprecedented pace, and astronomy is no exception to this trend. Availability of larger datasets may appear helpful for more effective decision making, but this is not so when this increase is primarily in data dimensionality. 
A large number of features can increase the noise of the data and thus the error of a learning algorithm. Feature selection is a solution for such problems. It reduces data dimensionality by removing irrelevant and redundant features. Besides maximizing model performance, other benefits include the ability to build simpler and faster models using only a subset of all features, as well as gaining a better understanding of the processes described by the data, by focusing on a selected subset of features.

In this paper, we deal with the issue of stellar characterization by spectral analysis, by employing a novel feature selection technique to automatically select wavelength points best suited for determination of the output parameters (T$_{\rm eff}$, $\log g$, and [Fe/H]). 
Using synthetic PHOENIX spectra, we show that the proposed algorithm is able to improve parameter prediction accuracy, in addition to being able to robustly select the most important absorption lines/ wavelength points in the wavelength range considered. This method is constructed in a modular fashion, and can be generalized to any regression task, within and outside of astronomy. At the time of writing this document, there exists only one astronomical publication dealing with feature selection \citep{d2018return}, and as such, we believe that the proposed method is an important addition to the literature.


\section{Feature Selection Methods} \label{sec:fs}
Feature selection techniques can be divided into three categories, depending on how they interact with the predictor -- `filter,' `wrapper,' and `embedded' \citep[see][for reviews]{guyon2003introduction, saeys2007review}. Filters operate directly on the dataset and select subsets of features as a pre-processing step, independently of the chosen predictor. 
Wrappers, on the other hand, select a subset of features based on the output of the prediction model. 
Embedded methods work similarly to wrappers, but use internal information from the prediction model to do feature selection \citep{saeys2008robust}. They often provide a good trade-off between performance and computational cost.
More often than not, different feature selection algorithms will choose different feature subsets \citep{saeys2008robust}. 
Ideally, we want any feature selection process not only to pick features with the best predictive power, but to also be robust -- small changes in the input dataset, or different runs of the feature selection model, should not affect the selected features. Robustness of feature selection processes has received relatively little attention, and most work has rather focused on the stability of single-feature selection techniques \citep{kvrivzek2007improving, saeys2008robust, raudys2006feature}. In this work, we explore the use of ensemble learning-based feature selection \citep{dietterich2000ensemble} to yield a stable set of selected features (wavelength points), while also using an ensemble of regressors to better predict the output variables. To the best of our knowledge, the current work is the first such work.

\section{Methodology}
\subsection{Data}
We use absorption spectra from the synthetic PHOENIX library
\citep{husser2013new} to test our proposed algorithm. We select a total of 1800 spectra as follows: T$_{\rm eff}
= 3500$ to $7000$ K in steps of 100 K, $\log g = 2.5$ to $5.0$ dex in steps of 0.5 dex, and [Fe/H] = --1 to +1 dex in steps of 0.25 dex. 
The spectra were convolved from the native resolution of 500,000 down to 10000, and uniformly re-sampled onto a wavelength range of 5000 -- 5450 \angstrom. This was done to match the operating characteristics of MARVELS \citep{ge2008multi} (a low-medium resolution spectrograph
commissioned as part of the Sloan Digital Sky Survey--III), since we plan to apply the proposed method to re-characterize stars observed with this instrument and compare the predicted parameter values to published ones. Finally, we picked 10$\%$ of all spectra (i.e., 180), distributed uniformly in the three--dimensional parameter space, as the training set, with the remaining 90$\%$ set aside as the test set.

\subsection{Ensemble Feature Selection}
Our goal is to formulate a feature selection strategy that not only reduces prediction error, but is also robust -- both to variations in the training dataset, and to the initial conditions of the machine learning algorithms
employed. Since features here refer to wavelength points or pixels, which have astrophysical origin and hence physical significance, it is imperative that any feature selection algorithm pick the same set of features over multiple runs. We achieve these goals using `ensembling' \citep{dietterich2000ensemble}, i.e., combining different ML models to overcome their individual limitations. Specifically, we use the techniques of `bagging' (short for `bootstrapped aggregation'), and `stacking' (combining the predictions, i.e., outputs, of one ML models as the input to another).

Using the training set of 180 stellar spectra, we create 1000 different `bags' of data, each of which contains approximately 67 ($\sqrt{4500}$, see \citet {friedman2001elements} for why we choose square root) features that are picked randomly (and with replacement across different `bags') from the input 4500 wavelength points. 
For each of these `bags', we create 100 bootstrapped versions (picking randomly and with replacement) of the training data (i.e., stellar spectra), and associate with them all a k-Nearest Neighbor regressor \citep{dudani1976distance} with distance metric |x-y|. For each `bag' we find the best k value by minimizing the `out-of-bag' error as follows -- we train the kNN on each of the 100 boostrapped datasets, predict the output variable for stellar samples left out during bootstrapping, and average the rms error with respect to the ground truth values across all 100 datasets. We choose the value of k that minimizes this error as the final value for that `bag'. We rank the input features according to their average errors across all `bags', discard the bottom $10\%$, and refit using three tree-based techniques: random forest, ada-boost, and extra trees regressor. We repeat this procedure of recursive backward elimination is until the rms stops decreasing. The entire process is also repeated separately for each of the three parameters of interest.



\begin{figure}[hbt!]
    \centering
    \includegraphics[width=1.0\textwidth]{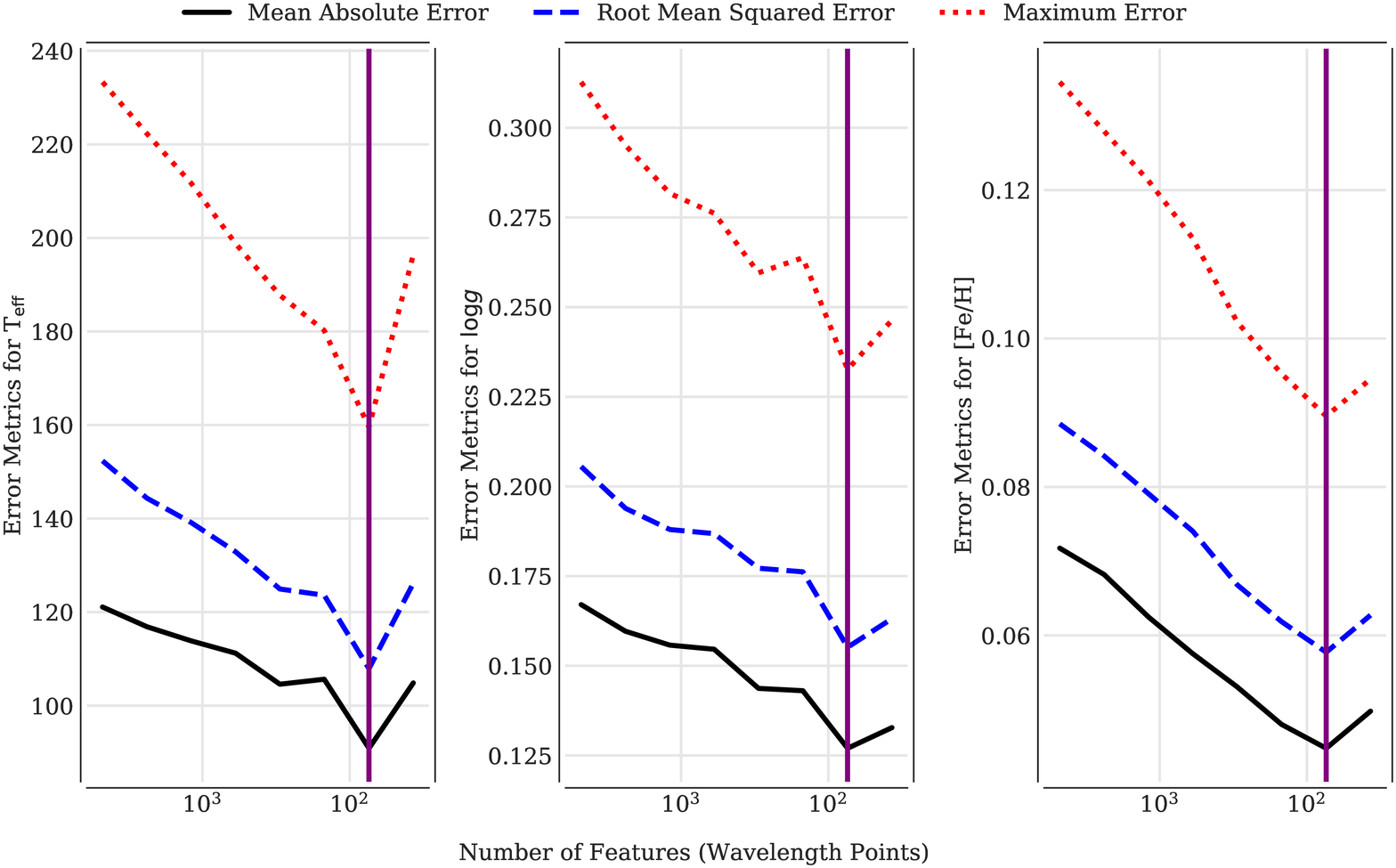}
    \caption{Error metrics as a function of number of features for T$_{\rm eff}$, $\log g$, and [Fe/H]. We plot three different error metrics between ground-truth values and the respective ensemble-predicted values. As redundant features are removed, prediction errors decrease (moving from left to right). All metrics start increasing again after the inflection point, when the actually informative features start being trimmed. The x-coordinate of this inflection-point is the optimal number of features for predicting the respective parameters.}
    \label{fig:fig1}
\end{figure}

\section{Results and Conclusions}
We have proposed an ensemble-based feature selection algorithm for dealing with datasets with large number of features and small number of samples. We have demonstrated its efficacy by successfully characterizing synthetic PHOENIX spectra (with predicted variables being T$_{\rm eff}$, $\log g$ and [Fe/H]) with 180 training samples and 4500 wavelength points (features) spanning the wavelength range from $\lambda = 5000~\angstrom$ to 5450 \angstrom. We were able to successfully select approximately 100 unique wavelength points and improve prediction accuracy for all three variables at the same time. The proposed method uses an ensemble of k-Nearest Neighbors to robustly select features in a recursive backward elimination procedure, and another ensemble of predictors to actually predict the output variables. Figure \ref{fig:fig1} clearly illustrates the decrease in various error metrics as the redundant wavelengths (features) are eliminated using our proposed backward elimination method.

In the future, we plan to move to a probabilistic framework for
feature selection, by using quantile regression as opposed to obtaining point estimates from our ensemble of predictors. This would naturally output an error range for all predicted variables rather than a single cross-validation error value. The method proposed here can, in principle, be applied to the calibration sample of any spectrograph to select wavelengths most suitable for parameter prediction; this would account for any instrument peculiarities in addition to capturing the relevant physics via the selected wavelengths. The features selected
in this manner can then be used to parameterize any new observed star, provided it lies in the original parameter space. We also plan to explore the performance of the proposed feature selection method as a function of input signal-to-noise ratio.

\bibliography{O1-3}

\end{document}